\documentstyle[psfig]{mn}

\def\etal{{\it et al.\ }}
\def\eg{{\it e.g.\ }}

\def\ie{{\it i.e.\ }}

\def\spose#1{\hbox to 0pt{#1\hss}}
\def\approxlt{\mathrel{\spose{\lower 3pt\hbox{$\sim$}}
	\raise 2.0pt\hbox{$<$}}}
\def\approxgt{\mathrel{\spose{\lower 3pt\hbox{$\sim$}}
	\raise 2.0pt\hbox{$>$}}}
\def\approxpropto{\mathrel{\spose{\lower 3pt\hbox{$\sim$}}
	\raise 2.0pt\hbox{$\propto$}}}
\mathchardef\twiddle="2218

\def\multleft#1{\hbox to size{\vbox {\halign {\lft{##}\cr #1}}\hfill}\par}
\def\multright#1{\hbox to size{\vbox {\halign {\rt{##}\cr #1}}\hfill}\par}

\def\today{\ifcase\month\or January\or February\or March\or April\or May\or
      June\or July\or August\or September\or October\or November\or December\fi
      \space\number\day, \number\year}
\def\<{\thinspace}

\def\erg{{\rm\thinspace erg}}

\def\km{{\rm\thinspace km}}

\def\Mpc{{\rm\thinspace Mpc}}

\def\s{{\rm\thinspace s}}

\def\ergps{\hbox{$\erg\s^{-1}\,$}}

\def\kmps{\hbox{$\km\s^{-1}\,$}}

\def\kmpspMpc{\hbox{$\kmps\Mpc^{-1}$}}

\title[Cooling flows and the $T_{\rm X}-L_{\rm Bol}$ relation]
{The impact of cooling flows on the $T_{\rm X}-L_{\rm Bol}$ relation 
for the most luminous clusters. }
\author[S.W. Allen and A.C. Fabian]
{\parbox[]{6.in} {S.W. Allen and A.C. Fabian\\
\footnotesize
Institute of Astronomy, Madingley Road, Cambridge CB3 OHA\\
}}
\begin{document}
\date{Submitted 1997 December 3}
\maketitle
\begin{abstract}
We examine the effects of cooling flows on the $T_{\rm X}-L_{\rm Bol}$
relation for a sample of the most X-ray luminous ($L_{\rm Bol} > 10^{45}$
\ergps) clusters of galaxies known. 
Using high-quality ASCA X-ray spectra and ROSAT images 
we explicitly account for the effects of cooling flows on the X-ray 
properties of the clusters and show that 
this reduces the previously-noted dispersion in the $T_{\rm X}-L_{\rm
Bol}$ relationship. More importantly, the   
slope of the relationship is flattened 
from $L_{\rm Bol} \propto T_{\rm X}^3$
to approximately $L_{\rm Bol} \propto T_{\rm X}^2$, in agreement with 
recent theoretical models which include the effects of shocks 
and preheating on the X-ray gas. We find no evidence for evolution in the 
$T_{\rm X}-L_{\rm Bol}$ relation within $z \sim 0.3$. Our results 
demonstrate that the effects of cooling flows must be accounted for before 
cosmological parameters can be determined from X-ray observations of 
clusters. The results presented here should provide a reliable basis for 
modelling the $T_{\rm X}-L_{\rm Bol}$ relation at high X-ray 
luminosities. 

\end{abstract}


\section{Introduction}

The baryonic content of clusters of galaxies is dominated by the 
metal-rich, X-ray luminous intracluster medium (ICM) that pervades
the cluster potentials. X-ray observations permit precise 
measurements of the X-ray luminosities and temperatures 
($T_{\rm X}$) of clusters, 
which can be related to the masses of these systems. 
For self-similar clusters with a characteristic density that
scales as the mean density of the universe, 
$T_{\rm X} \propto M^{2/3}(1+z_f)$, where $z_f$ is the redshift of formation of 
the cluster (e.g. Kaiser 1991; Evrard \& Henry 1991).  Since 
the bolometric luminosity is dominated by bremsstrahlung emission 
at X-ray wavelengths, we can then show that
$L_{\rm Bol} \propto f^2 T_{\rm X}^2 (1+z_f)^{3/2}$, where $f$ is the 
mass fraction in X-ray gas. (The redshift dependency holds for 
an Einstein-de-Sitter universe and is modified for lower values of
$\Omega_0$; \eg Eke, Navarro \& Frenk 1997).  
In principle, studies of the $L_{\rm Bol}-T_{\rm X}$ relation may 
be used to examine the evolution of clusters and should, for example, 
enable the range of cluster formation redshifts to be estimated 
(\eg Scharf \& Mushotzky 1997).

It is well known that the observed $L_{\rm Bol}-T_{\rm X}$ relation for
clusters contains a significant intrinsic dispersion. 
Within the simple theoretical framework outlined above, this 
dispersion  should relate to the ages of the clusters. However, 
such simple models do not account for 
important physical processes that could effect the
X-ray properties of clusters and, in particular, the effects 
of radiative cooling.

In the central regions of most ($70-90$ per cent) clusters, the cooling 
time of the ICM is significantly less than the Hubble time 
(Edge \etal 1992; Peres \etal 1997), which leads to the 
formation of cooling flows (Fabian 1994; Allen \& Fabian 1997). 
Fabian \etal (1994) showed that clusters with cooling flows are offset in 
the ($2-10$ keV) $L_{\rm X}-T_{\rm X}$ plane such that they have lower temperatures for 
a given luminosity (or alternatively higher luminosities for a given 
temperature). Before concluding that cooling-flow clusters are simply older 
than non-cooling flow systems, however, one must account for 
the effects of the cooling gas on the X-ray properties of the clusters. 
Cooling flows can account for up to 70 per cent of the total X-ray
luminosity of a cluster (Peres \etal 1997; Allen \etal 1998) and 
contain gas with a wide range of densities and 
temperatures. A spectral fit to a cooling-flow cluster with a simple 
isothermal model will yield only a  mean emission-weighted 
temperature that will be reduced with respect to the virial value.

Evidence for this effect was presented by 
Allen, Fabian \& Kneib (1996), from a study of the nearby, 
luminous cooling-flow cluster PKS0745-191. These authors showed that 
explicitly accounting for the effects of 
the large cooling flow in this cluster leads to an X-ray determined mass value in 
excellent agreement with 
gravitational lensing measurements, but as much as a factor 
3 greater than the mass inferred from a simple 
single-temperature X-ray analysis. 
Allen (1998) presented results for a larger sample of
clusters and showed that 
for all of the cooling flow clusters in that sample  
the virial temperatures, determined from multiphase X-ray analyses 
which accounted for the effects of the cooling flows, 
were significantly higher than those inferred from 
simple single-temperature models and lead to consistent 
X-ray and lensing mass measurements. 

A second issue relating to the $L_{\rm Bol}-T_{\rm X}$ 
results is the slope of the relation, which appears to fit 
$L_{\rm Bol}\propto T_{\rm X}^3$ (\eg Edge \& Stewart 1991; David \etal 1993; 
Fabian \etal 1994) rather than $L_{\rm Bol}\propto 
T_{\rm X}^2$, as would be expected for simple gravitational collapse. 
This again suggests the absence of important physics from 
the models. One well-discussed possibility is
that of pre-heating (Kaiser 1991; Evrard \& Henry 1991) in which 
energy sources such as stars, supernovae and AGN pre-heat the gas 
in the subclumps that merge to form present-day clusters. More recently, 
Cavaliere \etal (1997) have shown that accounting for 
the effects of shocks in cluster formation (with plausible 
levels of pre-heating) can lead  to $L_{\rm Bol}\propto 
T_{\rm X}^3$ at intermediate X-ray
luminosities, although at the highest luminosities 
the relation should flatten to 
$L_{\rm Bol}\propto T_{\rm X}^2$. 

In this Letter we use the results from multiphase analyses 
of ASCA spectra and ROSAT High Resolution Imager (HRI) data, for a 
sample of thirty of the most X-ray luminous 
known clusters (with $L_{\rm Bol} > 10^{45}$ \ergps), 
to examine the effects of cooling flows on the 
$L_{\rm Bol}-T_{\rm X}$ relation. We show that explicitly 
accounting for the effects of cooling flows leads to 
a significant flattening of the slope of the relation, 
to a value $L_{\rm Bol} \approxpropto 
T_{\rm X}^2$, in agreement with the theoretical
predictions. Throughout this Letter, we assume 
$H_0$=50 \kmpspMpc, $\Omega = 1$ and
$\Lambda = 0$.

\section{Observations and data analysis}

\begin{table*}
\vskip 0.2truein
\begin{center}
\caption{Bolometric luminosities, temperatures and X-ray gas mass
fractions} 
\vskip 0.2truein
\begin{tabular}{ c c c c c c c c c c c c c }
\hline                                                                                                                               
\multicolumn{1}{c}{} &
\multicolumn{1}{c}{} &
\multicolumn{1}{c}{} &
\multicolumn{1}{c}{} &
\multicolumn{4}{c}{LUMINOSITY} &
\multicolumn{1}{c}{} &
\multicolumn{2}{c}{TEMPERATURE} &
\multicolumn{1}{c}{} &
\multicolumn{1}{c}{$M_{\rm gas}/M_{\rm total}$ } \\
\multicolumn{1}{c}{} &
\multicolumn{1}{c}{} &
\multicolumn{1}{c}{} &
\multicolumn{1}{c}{} &
\multicolumn{4}{c}{($10^{44}$ \ergps)} &
\multicolumn{1}{c}{} &
\multicolumn{2}{c}{(keV)} &
\multicolumn{1}{c}{} &
\multicolumn{1}{c}{(at 500 kpc)} \\
\hline                                                                                                                               
&&&&&&&& \\                                                                                                                         
\multicolumn{1}{c}{COOLING FLOWS} &
\multicolumn{1}{c}{} &
\multicolumn{1}{c}{$z$} &
\multicolumn{1}{c}{} &
\multicolumn{2}{c}{MODEL A} &
\multicolumn{2}{c}{MODEL C} &
\multicolumn{1}{c}{} &
\multicolumn{1}{c}{MODEL A} &
\multicolumn{1}{c}{MODEL C} &
\multicolumn{1}{c}{} &
\multicolumn{1}{c}{MODEL C} \\
&&&&($L_{\rm X}$)&($L_{\rm Bol}$)&($L_{\rm X}$)&($L_{\rm Bol}$)&&($kT_{\rm X}$)&($kT_{\rm X}$)&&($f$)\\
  
&&&&&&&& \\                                                                                                                         
Abell 478       & ~ &  0.088 & ~ & 24.2  & 53.8  &  24.2 & 63.6   & & $6.02^{+0.11}_{-0.11}$  &  $8.1^{+1.2}_{-0.8}$    && $0.22^{+0.03}_{-0.03}$   \\  
Abell 586       & ~ &  0.171 & ~ & 10.7  & 22.7  &  10.7 & 26.1   & & $7.15^{+0.72}_{-0.62}$  &  $10.7^{+10.3}_{-3.9}$  && $0.10^{+0.06}_{-0.05}$   \\  
PKS0745-191     & ~ &  0.103 & ~ & 29.5  & 65.2  &  29.5 & 74.0   & & $6.71^{+0.14}_{-0.14}$  &  $8.7^{+1.6}_{-1.2}$    && $0.20^{+0.03}_{-0.03}$   \\  
IRAS09104+4109  & ~ &  0.442 & ~ & 22.0  & 46.3  &  21.8 & 55.1   & & $6.28^{+0.55}_{-0.48}$  &  $8.5^{+5.6}_{-1.8}$    && $0.12^{+0.03}_{-0.05}$   \\  
Abell 963       & ~ &  0.206 & ~ & 12.7  & 27.0  &  12.7 & 27.0   & & $6.16^{+0.34}_{-0.31}$  &  $6.13^{+0.45}_{-0.30}$ && $0.21^{+0.03}_{-0.02}$   \\  
Zwicky 3146     & ~ &  0.291 & ~ & 36.9  & 78.6  &  37.3 & 91.8   & & $6.41^{+0.26}_{-0.25}$  &  $11.3^{+5.8}_{-2.7}$   && $0.16^{+0.05}_{-0.05}$   \\  
Abell 1068      & ~ &  0.139 & ~ & 7.00  & 16.1  &  6.80 & 21.8   & & $4.21^{+0.18}_{-0.14}$  &  $5.5^{+1.4}_{-0.9}$    && $0.17^{+0.04}_{-0.03}$   \\  
Abell 1413$^*$  & ~ &  0.143 & ~ & 15.5  & 32.9  &  15.4 & 36.4   & & $7.54^{+0.29}_{-0.27}$  &  $8.5^{+1.3}_{-0.8}$    && $0.10^{+0.01}_{-0.01}$   \\  
Abell 1689      & ~ &  0.184 & ~ & 32.4  & 70.3  &  32.2 & 62.6   & & $9.66^{+0.37}_{-0.34}$  &  $10.0^{+1.2}_{-0.8}$   && $0.18^{+0.01}_{-0.02}$   \\  
Abell 1704      & ~ &  0.216 & ~ & 8.32  & 18.6  &  8.19 & 22.9   & & $4.73^{+0.38}_{-0.33}$  &  $5.7^{+3.5}_{-1.3}$    && $0.21^{+0.06}_{-0.08}$   \\  
RXJ1347.5-1145  & ~ &  0.451 & ~ & 94.4  & 217.8 &  93.5 & 225.0  & & $12.46^{+0.90}_{-0.82}$ &  $26.4^{+7.8}_{-12.3}$  && $0.09^{+0.06}_{-0.02}$   \\  
Abell 1795      & ~ &  0.063 & ~ & 11.0  & 23.6  &  11.0 & 25.2   & & $5.40^{+0.08}_{-0.09}$  &  $5.87^{+0.29}_{-0.25}$ && $0.23^{+0.01}_{-0.01}$   \\  
MS1358.4+6245   & ~ &  0.327 & ~ & 10.8  & 22.8  &  10.5 & 26.3   & & $7.48^{+0.83}_{-0.70}$  &  $7.5^{+7.1}_{-1.5}$    && $0.17^{+0.06}_{-0.08}$   \\  
Abell 1835      & ~ &  0.252 & ~ & 45.4  & 96.3  &  44.9 & 106.2  & & $8.21^{+0.31}_{-0.29}$  &  $9.8^{+2.3}_{-1.3}$    && $0.24^{+0.03}_{-0.04}$   \\  
MS1455.0+2232   & ~ &  0.258 & ~ & 16.6  & 37.0  &  16.4 & 44.7   & & $4.83^{+0.22}_{-0.21}$  &  $5.6^{+3.1}_{-0.9}$    && $0.26^{+0.05}_{-0.09}$   \\  
Abell 2029      & ~ &  0.077 & ~ & 19.9  & 41.8  &  19.9 & 45.5   & & $7.72^{+0.13}_{-0.13}$  &  $8.47^{+0.41}_{-0.36}$ && $0.18^{+0.01}_{-0.02}$   \\  
Abell 2142      & ~ &  0.089 & ~ & 30.4  & 64.6  &  30.4 & 67.8   & & $8.67^{+0.32}_{-0.30}$  &  $9.3^{+1.3}_{-0.7}$    && $0.18^{+0.01}_{-0.03}$   \\  
Abell 2204      & ~ &  0.152 & ~ & 35.2  & 74.2  &  34.6 & 85.5   & & $7.40^{+0.30}_{-0.28}$  &  $9.2^{+2.5}_{-1.1}$    && $0.20^{+0.02}_{-0.04}$   \\  
Abell 2261      & ~ &  0.224 & ~ & 23.9  & 51.3  &  23.3 & 58.3   & & $8.82^{+0.61}_{-0.54}$  &  $10.9^{+5.9}_{-2.2}$   && $0.13^{+0.04}_{-0.04}$   \\  
MS2137.3-2353   & ~ &  0.313 & ~ & 17.0  & 36.9  &  16.6 & 44.1   & & $5.16^{+0.39}_{-0.34}$  &  $5.2^{+1.8}_{-0.7}$    && $0.21^{+0.03}_{-0.05}$   \\  
Abell 2390      & ~ &  0.233 & ~ & 41.3  & 90.8  &  41.0 & 101.1  & & $10.13^{+1.22}_{-0.99}$ &  $14.5^{+15.5}_{-5.2}$  && $0.15^{+0.07}_{-0.08}$   \\  
&&&&&&&& \\                                                                                                                             
&&&&&&&& \\                                                                                                                             
\hline                                                                                                                               
&&&&&&&& \\                                                                                                                             
\multicolumn{1}{c}{NON-COOLING FLOWS} &											      
\multicolumn{1}{c}{} &													      
\multicolumn{1}{c}{$z$} &
\multicolumn{1}{c}{} &
\multicolumn{2}{c}{MODEL A} &
\multicolumn{2}{c}{} &
\multicolumn{1}{c}{} &
\multicolumn{1}{c}{MODEL A} &
\multicolumn{1}{c}{} &
\multicolumn{1}{c}{} &
\multicolumn{1}{c}{MODEL A} \\
&&&&($L_{\rm X}$)&($L_{\rm Bol}$)&&&&($kT_{\rm X}$)&&&($f$)\\
&&&&&&&& \\
Abell 2744      & ~ &  0.308 & ~ & 31.3  & 70.1   & --- & --- & & $11.04^{+0.79}_{-0.73}$ & --- & & $0.15^{+0.02}_{-0.01}$ \\
Abell 520       & ~ &  0.203 & ~ & 15.8  & 34.1   & --- & --- & & $8.33^{+0.76}_{-0.67}$  & --- & & $0.15^{+0.03}_{-0.02}$ \\  
Abell 665       & ~ &  0.182 & ~ & 17.8  & 38.5   & --- & --- & & $9.03^{+0.58}_{-0.52}$  & --- & & $0.12^{+0.01}_{-0.01}$ \\  
Abell 773       & ~ &  0.217 & ~ & 14.8  & 32.0   & --- & --- & & $9.29^{+0.69}_{-0.60}$  & --- & & $0.11^{+0.02}_{-0.02}$ \\  
Abell 2163      & ~ &  0.208 & ~ & 61.0  & 147.0  & --- & --- & & $13.83^{+0.78}_{-0.74}$ & --- & & $0.13^{+0.01}_{-0.01}$ \\  
Abell 2218      & ~ &  0.175 & ~ & 10.8  & 23.1   & --- & --- & & $7.05^{+0.36}_{-0.35}$  & --- & & $0.14^{+0.02}_{-0.02}$ \\  
Abell 2219      & ~ &  0.228 & ~ & 38.9  & 89.4   & --- & --- & & $12.42^{+0.77}_{-0.69}$ & --- & & $0.15^{+0.01}_{-0.02}$ \\  
Abell 2319      & ~ &  0.056 & ~ & 16.9  & 36.8   & --- & --- & & $9.30^{+0.41}_{-0.38}$  & --- & & $0.12^{+0.01}_{-0.01}$ \\  
AC114           & ~ &  0.312 & ~ & 17.5  & 38.1   & --- & --- & & $9.76^{+1.04}_{-0.85}$  & --- & & $0.10^{+0.01}_{-0.02}$ \\  
&&&&&&&& \\                                                                                                                         
\hline                                                                                                                                                   
&&&&&&&& \\                                                                                                                         
\end{tabular}
\end{center}
\parbox {7in}
{Notes: From left to right we list the redshifts ($z$), $2-10$ keV X-ray 
luminosities ($L_{\rm X}$), bolometric
luminosities ($L_{\rm Bol}$: corrected for the effects of Galactic and
intrinsic absorption), ambient cluster temperatures ($kT_{\rm X}$) and 
X-ray gas mass fractions ($f$) determined at a radius of 500 kpc. 
Errors are the 90 per cent ($\Delta \chi^2 = 2.71$) confidence limits 
on a single interesting parameter.}
\end{table*}

Full discussions of the reduction and analysis of the X-ray data and the
properties of the cluster sample are presented by Allen \etal (1998). 
The ROSAT HRI data were used 
to map the X-ray surface brightness profiles of the clusters 
and to determine whether the individual clusters have cooling flows or not. 
We define cooling-flow (CF) clusters to be 
those systems for which the upper (90 per cent confidence) limit to
the central cooling time, determined from the HRI data, is
less than $10^{10}$ yr. Non-cooling flow (NCF) clusters 
are those systems with upper limits to their central cooling times $>
10^{10}$ yr. Using this simple classification we identify 21 CF
and 9 NCF clusters in the sample, with a mean redshift for
both subsamples of $\bar{z} = 0.21$. 

The ASCA spectra were fitted with a series of appropriate spectral models 
from which the bolometric luminosity ($L_{\rm Bol}$), $2-10$ keV X-ray
luminosity ($L_{\rm X}$), 
and temperature ($kT_{\rm X}$) 
measurements were made. Spectral model A consisted of an isothermal
plasma in collisional equilibrium, 
at the optically-determined redshift for the 
cluster, and absorbed by the nominal Galactic column density 
(Dickey \& Lockman 1990). 
The free parameters in this model were the temperature ($kT_{\rm X}$) and
metallicity ($Z$; determined relative to the solar values of 
Anders \& Grevesse 1989) of the plasma and the emission measures in the four 
detectors. Secondly, model B, which was identical to model A but with the 
absorbing column density ($N_{\rm H}$) also included as a free parameter in 
the fits. Thirdly, model C, which included an additional component
accounting for the emission from the cooling flows in the clusters (Johnstone \etal 1992). The normalization of the cooling-flow 
component was parameterized in terms of a mass deposition rate, ${\dot M}$, 
which was free parameter in the fits. The cooling flows were also assumed to 
be absorbed by an intrinsic column density, $\Delta N_{\rm H}$ (Allen \&
Fabian 1997 and references therein), which was 
a further free parameter. The results on the mass deposition
rates and intrinsic column densities are presented by Allen
\etal (1998). The relationships between cooling flows and
metallicity measurements for clusters are discussed in a companion Letter 
(Allen \& Fabian 1998).

The modelling of the ASCA spectra was carried out using the XSPEC spectral
fitting package (version 9.0; Arnaud 1996). The spectra were modelled
using the plasma codes of Kaastra \& Mewe (1993; incorporating the Fe L
calculations by Liedhal in XSPEC version 9.0) 
and the photoelectric absorption models of Balucinska-Church \&
McCammon (1992). The data from all four ASCA detectors were analysed 
simultaneously with the fit parameters linked to take the same
values across the data sets. The exceptions were the emission
measures of the ambient cluster gas in the four detectors which, due to 
the different extraction radii used, were allowed to fit independently.
The luminosity and temperature results for the clusters, and the X-ray
gas mass fractions at a radius of 500 kpc (determined from the HRI data
using the ASCA constraints) are summarized in Table 1. The $L_{\rm Bol}$ values 
are measured in the GIS3 detector.

\section{The $T_{\rm X}-L_{\rm Bol}$ relation}

\begin{table*}
\vskip 0.2truein
\begin{center}
\caption{The power-law fits to the $T_{\rm X} - L_{\rm Bol}$ 
data}
\vskip 0.2truein
\begin{tabular}{ c c c c c c c c c }
 \hline                                                                               
\multicolumn{1}{c}{} &
\multicolumn{1}{c}{} &
\multicolumn{1}{c}{} &
\multicolumn{2}{c}{MODEL A FOR CF CLUSTERS} &
\multicolumn{1}{c}{} &
\multicolumn{1}{c}{} &
\multicolumn{2}{c}{MODEL C FOR CF CLUSTERS} \\
\multicolumn{1}{c}{} &
\multicolumn{1}{c}{} &
\multicolumn{1}{c}{} &
\multicolumn{1}{c}{$\chi^2$} &
\multicolumn{1}{c}{BCES} &
\multicolumn{1}{c}{} &
\multicolumn{1}{c}{} &
\multicolumn{1}{c}{$\chi^2$} &
\multicolumn{1}{c}{BCES} \\
 \hline                                                                               	                                                               
&&&&&&&& \\                                                                         	                                                               
                   &&  Q & $0.296 \pm 0.014$  & $0.325 \pm 0.061$  & ~~~~~~ & Q & $0.423 \pm 0.028$ & $0.429 \pm 0.079$ \\    
ALL CLUSTERS       &&  P & $2.15 \pm 0.12$    & $2.22 \pm 0.54$    & ~~~~~~ & P & $1.72 \pm 0.19$   & $1.66 \pm 0.52$  \\ 
                   && $\chi^2$/DOF & 1744/28  & ---                & ~~~~~~ & $\chi^2$/DOF  & 178.6/28        & ---    \\
&&&&&&&& \\                                         	                                                                                                                                   
                   &&  Q & $0.284 \pm 0.015$  & $0.311 \pm 0.061$  & ~~~~~~ & Q & $0.450 \pm 0.058$ & $0.470 \pm 0.109$  \\   
CFs ONLY           &&  P & $2.19 \pm 0.13$    & $2.11 \pm 0.51$    & ~~~~~~ & P & $1.41 \pm 0.30$   & $1.34 \pm 0.58$   \\  
                   && $\chi^2$/DOF  & 1062/19 & ---                & ~~~~~~ & $\chi^2$/DOF & 38.5/19  & ---      \\
&&&&&&&& \\                                         	                                                                                                                                   
                   &&  Q & $0.345 \pm 0.033$  & $0.343 \pm 0.053$  & ~~~~~~ & Q &            ---   & ---      \\  
NCFs ONLY          &&  P & $2.57 \pm 0.33$    & $2.62 \pm 0.53$    & ~~~~~~ & P &            ---   & ---      \\       
                   &&  $\chi^2$/DOF & 16.8/7  & ---                & ~~~~~~ & $\chi^2$/DOF & ---   & ---      \\
&&&&&&&& \\
\hline 
&&&&&&&& \\                                                                         	                                                               
\end{tabular}
\end{center}
\parbox {7in}
{The results from fits to the $kT_{\rm X} - L_{\rm Bol}$ data with 
power-law models of the form $kT_{\rm X} = P L_{\rm Bol}^Q$. 
$kT_{\rm X}$ is in units of keV and $P$ in units of $10^{44}$ \ergps. 
Both standard $\chi^2$ and the Akritas \& Bershady (1996) BCES
modification of the ordinary least squares estimator have been used. 
Error bars from the $\chi^2$ analysis are formal $\Delta \chi^2 =
2.71$ confidence limits. Errors on the BCES results are the standard
deviations determined by bootstrap resampling. Where the reduced $\chi^2$
values are large, the BCES estimator is more applicable and
these results should be used. [We have used the 
BCES$(X_2|X_1)$ estimator of Akritas \& Bershady (1996). The results 
obtained are in excellent agreement with those 
determined using the orthogonal distance BCES estimator. 
We note that use of the BCES$(X_1|X_2)$ estimator, in the fits to the
whole sample and subsample of CF clusters with spectral model A, leads to
somewhat steeper slopes, although the bootstrap errors are 
significantly ($3-6$ times) larger than those obtained with the other 
BCES estimators. Using spectral model C for the CF clusters we obtain
excellent agreement between the fit results for all BCES estimators.]
}
\end{table*}

\subsection{The dispersion in the relation}

\begin{figure}
\centerline{\hspace{0cm}\psfig{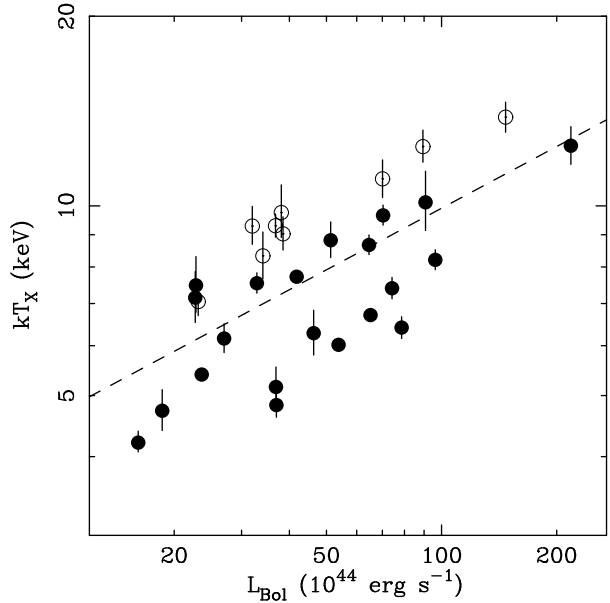}}
\caption{The $kT_{\rm X}-L_{\rm Bol}$ relationship determined 
using the isothermal spectral model (A) for all clusters. The dashed line is
the best-fitting power-law model using the BCES least-squares estimator
(Table 2). The CF and NCF systems are plotted as 
filled and open circles, respectively. The plotted error bars are the 
$\Delta \chi^2 =2.71$ confidence limits on the $kT$ values (Table 1). 
}
\end{figure}

\begin{figure}
\centerline{\hspace{0cm}\psfig{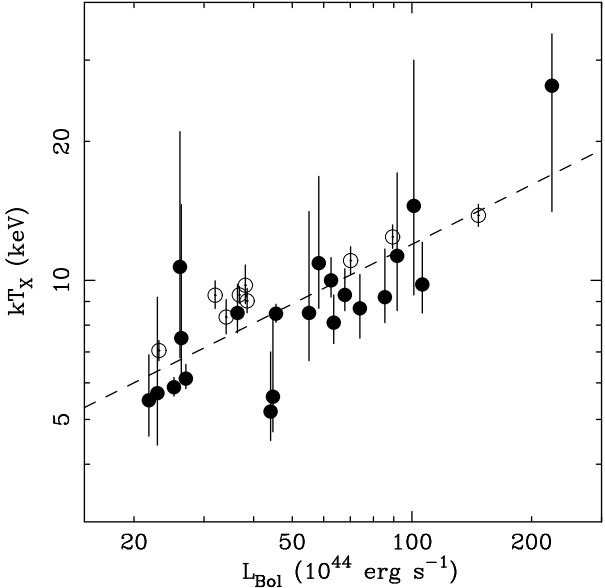}}
\caption{The $kT_{\rm X}-L_{\rm Bol}$ relationship determined 
using model C (which explicitly accounts for the cooling flows) 
for the CF  clusters and model A for the NCF systems. 
The dashed line is the best-fitting power law model 
using the BCES least-squares estimator.  Other details as in 
Fig. 1.}
\end{figure}

The bolometric luminosities and X-ray temperatures for the clusters
determined with the isothermal spectral model (A) are plotted in Fig.~1. 
$kT_{\rm X}$ has been adopted as the ordinate since this variable has 
the dominant statistical errors. (The statistical errors on the
luminosities are negligible.) The wide dispersion in the 
$kT_{\rm X}-L_{\rm Bol}$  plane is evident and is much greater than the 
average statistical error bar on the temperatures. The CF clusters clearly 
lie below the NCF objects.

We have fitted the $kT_{\rm X}-L_{\rm Bol}$ data with power law models of
the form $kT_{\rm X} = P L_{\rm Bol}^Q$, using both standard $\chi^2$ and
the Akritas \& Bershady (1996) modification of the 
ordinary least squares statistic (referred to by those authors
as bivariate correlated errors and intrinsic scatter or BCES). 
The results are summarized in Table 2.

With the temperatures and luminosities for the CF clusters 
corrected for the effects of cooling
flows (using spectral model C), 
the $T_{\rm X}-L_{\rm Bol}$ relation appears more uniform (Fig. 2). 
The mean weighted variance about the best-fitting BCES 
curves drops from 2.22 keV$^2$, using Model A for all clusters, 
to 1.09 keV$^2$, when using model C for the CF systems. 

A further significant reduction in the dispersion is obtained
by modelling the subsamples of CF and NCF clusters separately. 
For the CF clusters, the mean weighted variance about the best-fitting
(BCES) curve is 0.98 keV$^2$ using model A, and 0.46 keV$^2$ using model C, 
indicating a significant improvement to the power-law fit 
when accounting for the effects of the cooling flows. (This 
improvement is also reflected in the $\chi^2$ values listed in Table 2). 
For the subsample of NCF clusters, the mean weighted variance about the 
best-fitting curve is only 0.24 keV$^2$. 
(We note that at least part of the residual offset between the CF and 
NCF clusters in Fig. 2 can be attributed to the effects of 
subcluster merger events in the NCF clusters, which will disrupt their core 
regions and reduce the central X-ray gas densities and luminosities by up 
to a factor of $\sim 2$.)

\subsection{The intrinsic slope}

The most remarkable change to the $T_{\rm X}-L_{\rm Bol}$ results upon
correcting for the effects of the cooling flows 
is in the slope of the relation. Using the simple isothermal
model (A) for all clusters we obtain  $T_{\rm X} \propto
L_{\rm Bol}^{0.33}$ (using the BCES estimator),
in general agreement with the results from previous studies (\eg Edge \&
Stewart 1991; David \etal 1993; Fabian \etal 1994). Using spectral model C 
for the CF clusters, however, the best-fitting slope determined from a
fit to the whole sample is steepened to $T_{\rm X} \propto L_{\rm
Bol}^{0.43}$.

Examining the subsamples of CF and NCF clusters separately, we
obtain for the NCF clusters $T_{\rm X} \propto L_{\rm Bol}^{0.34}$ (BCES
using spectral model A), with a marginally 
acceptable value for the reduced $\chi^2$ of 16.8 for 7 degrees of freedom. 
The high-luminosity NCF clusters thus maintain
the slope determined from studies of lower-luminosity systems. 
For the CF clusters, using spectral model A we find 
$T_{\rm X} \propto L_{\rm Bol}^{0.31}$ (BCES),  
with an unacceptable $\chi^2$ of 1062 for 19 degrees of freedom. 
Using spectral model C for the CF clusters, however, the slope is steepened 
to $T_{\rm X} \propto L_{\rm Bol}^{0.47}$ 
and the $\chi^2$ value substantially improved to a marginally 
acceptable value of 38.5.

Our results demonstrate that accounting for the effects of 
cooling flows on the X-ray properties of the clusters steepens the
$T_{\rm X}-L_{\rm Bol}$ relation (or conversely 
flattens the $L_{\rm Bol}-T_{\rm X}$ relation). 
In particular, for the CF clusters the corrected 
$L_{\rm Bol}-T_{\rm X}$ relation is consistent with 
$L_{\rm Bol}\propto T_{\rm X}^2$, in agreement with the
simple models outlined in Section 1. We note 
that this result is not dominated by the extreme cluster 
RXJ1347.5-1145, which has the largest statistical uncertainty 
associated with its temperature and, therefore, a low weight in
the fits. We have searched for biases in the results by splitting 
the clusters into two subsamples, with bolometric luminosities less than 
and greater than $5\times 10^{45}$ \ergps~(using spectral model C 
for the CF clusters). The fit results for the two subsamples 
show excellent agreement with each other and with the results for the sample 
as a whole.

The slopes of the relations are not significantly 
altered if the $2-10$ keV X-ray luminosity 
rather than the bolometric luminosity is used.
(This is not surprising since the bolometric luminosity of 
a hot, X-ray luminous cluster is dominated by the contribution in the 
2-10 keV band.) We note, however, that the $T_{\rm X}-L_{\rm Bol}$
relation for the NCF clusters is further flattened (\ie the
$L_{\rm Bol}-T_{\rm X}$ relation is steepened) if spectral model
B rather than model A is used for these systems
\ie if the absorbing column density is included as a free
parameter in the isothermal spectral analysis.
[We have used model A, in preference to model B, for our analysis of the 
NCF clusters because it provides temperature results 
in better agreement (where such a comparison is possible) with the results 
from previous X-ray experiments which were less sensitive to the precise 
column density values ({\it c.f.} David \etal 1993). For a more detailed 
discussion of the merits of the individual spectral models see 
Allen \etal (1998).]  

A comparison of our results with those of David \etal 
(1993) shows that our best-fitting relation, 
accounting for the effects of the cooling flows, 
under-predicts the observed cluster temperatures 
at $L_{\rm Bol}\sim 10^{44}\ergps$ by $\sim 1$ keV. 
Although the $T_{\rm X}-L_{\rm Bol}$ values for the lower-luminosity
clusters will also be adjusted once the effects of their cooling flows 
have been accounted for (this will be presented in future work) 
the effects are unlikely to be sufficient to bring the
lower-luminosity data points into agreement with
the slope extrapolated from the high luminosity clusters. 
This implies that the $T_{\rm X}-L_{\rm Bol}$ relation is
not a simple power-law across all luminosities but flattens at lower
$L_{\rm Bol}$ values (\ie the $L_{\rm Bol}-T_{\rm X}$ relation steepens). 
Our results for clusters with 
$L_{\rm Bol} > 10^{45}\ergps$ are in good 
agreement with the predictions of Cavaliere \etal (1997) and support the
suggestion that pre-heating of the ICM significantly effects 
the X-ray properties of (only) lower-luminosity clusters. 

\begin{figure}
\centerline{\hspace{0cm}\psfig{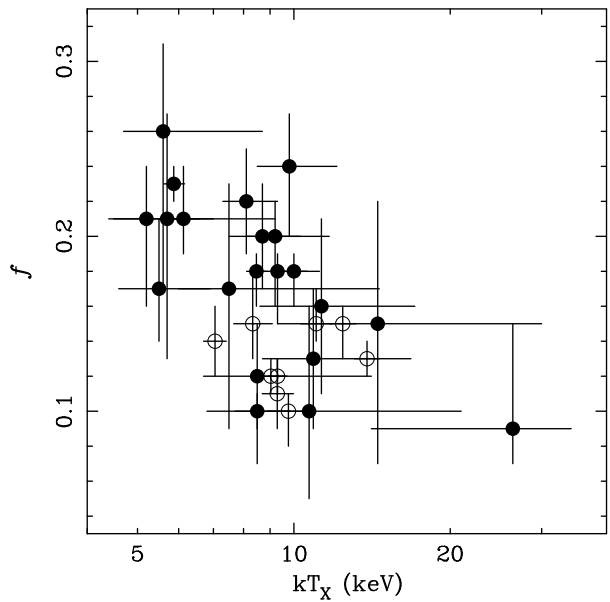}}
\caption{The X-ray gas mass fraction, $f$, 
measured at a radius of 500 kpc versus the X-ray temperature. 
} 
\end{figure}

Finally, we note that our results argue against the suggestion of 
David \etal (1993) that the $L_{\rm
Bol}\propto T_{\rm X}^3$ relation, determined from simple isothermal spectral
analyses, may be reconciled with $L_{\rm Bol}\propto T_{\rm X}^2$
if the X-ray gas mass fraction, $f$,  correlates with the X-ray 
temperature as $f \propto T_{\rm X}^{0.5}$. The results on $f$ and 
$kT_{\rm X}$ plotted in Fig.3 have a best-fitting power-law slope of
$f \propto T_{\rm X}^{-0.68\pm0.22}$ (using the BCES orthogonal distance
estimator), albeit with significant scatter.

\subsection{Evolution}

\begin{figure}
\centerline{\hspace{0cm}\psfig{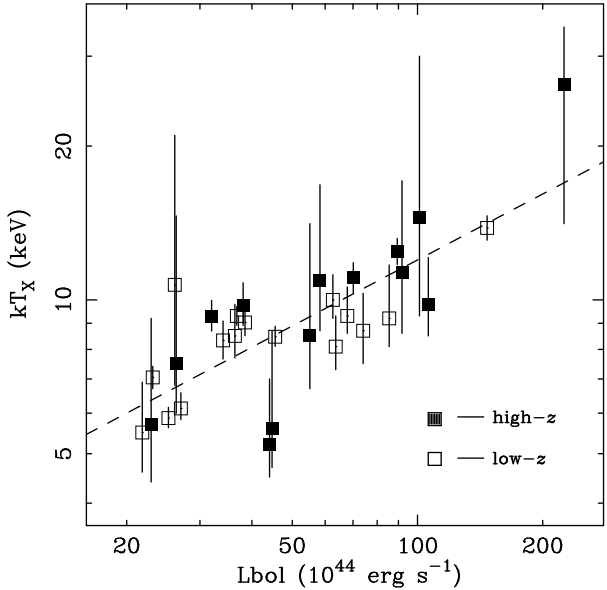}}
\caption{The $kT_{\rm X}-L_{\rm Bol}$ relationship separated into 
subsamples of low$-z$ and high$-z$ clusters (plotted as open and filled squares
respectively). The dashed line is the best-fitting (BCES) power-law model 
for the whole sample.}
\end{figure}

Within the simple theoretical framework outlined in Section 1, 
we may expect clusters that have formed earlier to have a lower 
normalization in the $T_{\rm X}-L_{\rm Bol}$ plane. 
We have examined the evidence for evolution in the $T_{\rm X}-L_{\rm
Bol}$ data by splitting the clusters into two subsamples with redshifts 
less and greater than the mean value of ${\bar z}=0.21$. 
This gives 16 clusters with
$z<0.21$ and 14 with $z>0.21$, with an approximately even mix of CF and
NCF systems in the two subsamples (5 and 4 NCFs
the low and high redshift subsamples, respectively).
The mean redshifts for the low$-z$ and high$-z$ clusters are 0.14 and 
0.29, respectively. 

Fig. 4 shows the $kT_{\rm X}-L_{\rm Bol}$ data for the two
subsamples (using spectral model C for the CF clusters). 
Using the low$-z$ clusters as a calibrator (Feigelson \& Babu 1992)
we determine an offset between the high$-z$ and low$-z$ systems such
that the normalization of the high$-z$ subsample is a
factor of $1.045 \pm 0.081$ higher than that for the low$-z$ clusters. 
We thus find no evidence for significant evolution in the 
$T_{\rm X}-L_{\rm Bol}$ relation over the range of redshifts covered by 
our study. 

The results presented in this Letter demonstrate that our understanding 
of the physics within clusters, and in particular the effects of
radiative cooling, must be improved before definitive statements 
about the origin and evolution of clusters 
can be made from the $T_{\rm X}-L_{\rm Bol}$ relation. 
NCF clusters generally show evidence for recent or ongoing
merger events, that have disrupted their central X-ray gas distributions 
and lead to apparent conflicts between X-ray and gravitational lensing mass
measurements for these systems (Allen 1998).
In contrast, optical, X-ray and gravitational lensing studies of CF
clusters show these systems to be regular and dynamically-relaxed 
and to exhibit excellent agreement between their X-ray and lensing mass 
measurements (Allen \etal 1996; Allen 1998). We conclude that 
the X-ray data for CF clusters, corrected for the
effects of cooling flows, should provide an
accurate measure of the virial properties of these clusters, 
and the best basis for comparisons with theoretical models.

\section*{Acknowledgments}

We thank M. Bershady and E. Feigelson for kindly providing copies
of their least-squares regression codes. We thank 
the Royal Society for support.


\begin{thebibliography}{}

\bibitem{} Allen S.W., 1998, MNRAS, in press (astro-ph/9710217)
\bibitem{} Allen S.W., Fabian A.C., 1997, MNRAS, 286, 583
\bibitem{} Allen S.W., Fabian A.C., 1998, MNRAS, submitted
\bibitem{} Allen S.W., Fabian A.C., Kneib J.-P., 1996, MNRAS, 279, 615
\bibitem{} Allen S.W. \etal, 1998, in preparation
\bibitem{} Akritas M.G., Bershady M.A., 1996, ApJ, 470, 706 
\bibitem{} Anders E., Grevesse N., 1989, Geochemica et Cosmochimica Acta 53, 197
\bibitem{} Arnaud K.A., 1996, in Astronomical Data Analysis Software and Systems V, eds. Jacoby G. and Barnes J., ASP Conf. Series volume 101, p17
\bibitem{} Balucinska-Church M., McCammon D., 1992, ApJ, 400, 699
\bibitem{} Dickey \& Lockman, 1990, Ann. Rev. Ast. Astr. 28, 215
\bibitem{} Cavaliere A., Menci N., Tozzi P., 1997, ApJ, 484, L21
\bibitem{} David L.P., Slyz A., Jones C., Forman W., Vrtilek S.D., Arnaud K.A., 1993, ApJ, 412, 479
\bibitem{} Dickey J.M., Lockman F.J., 1990, ARA\&A, 28, 215
\bibitem{} Edge A.C., Stewart A.C., 1991, MNRAS, 252, 414
\bibitem{} Edge A.C., Stewart G.C., Fabian A.C., 1992, MNRAS, 258, 177
\bibitem{} Eke V.R., Navarro J.F., Frenk C.S., 1997, ApJ, submitted (astro-ph/9708070)
\bibitem{} Evrard A.E., Henry J.P., 1991, ApJ, 383, 95
\bibitem{} Fabian A.C, 1994, A\&AR, 32, 277
\bibitem{} Fabian A.C., Crawford C.S., Edge A.C., Mushotzky R.F., 1994, MNRAS, 267, 779
\bibitem{} Feigelson E.D., Babu G.J., 1992, ApJ, 397, 55
\bibitem{} Johnstone R.M., Fabian A.C., Edge A.C., Thomas P.A., 1992, MNRAS, 255, 431
\bibitem{} Kaastra J.S., Mewe R., 1993, Legacy, 3, HEASARC, NASA
\bibitem{} Kaiser N., 1991, ApJ, 383, 104
\bibitem{} Peres C.B., Fabian A.C., Edge A.C., Allen S.W., Johnstone R.M., White D.A., 1997, MNRAS, in press
\bibitem{} Scharf C.A., Mushotzky R.F., 1997, ApJ, 485L, 65

\end{thebibliography}
\end{document}